\documentclass[twoside]{photon2007}
\usepackage[latin1]{inputenc}
\usepackage[dvips]{graphicx,epsfig,color}
\usepackage{wrapfig,rotating}
\usepackage{amssymb,amsmath,array}

\begin{document}
\title{
Minijets, soft gluon resummation and photon cross-sections}
\author{R.M. Godbole$^1$, A. Grau$^2$, G. Pancheri$^3$  and Y.N. Srivastava$^4$
\vspace{.3cm}\\
1- Centre for High Energy Physics, Indian Institute of Science,\\
 Bangalore, 560012, India.
\vspace{.1cm}\\
2-  Departamento de F\'\i sica Te\'orica y del Cosmos \\
Universidad de Granada, 18071 Granada, Spain
\vspace{.1cm}\\
3-INFN Frascati National Laboratories\\
Via Enrico Fermi 40, I-00044 Frascati, Italy
\vspace{.1cm}\\
4-INFN and Physics Department, University of Perugia \\
Via A. Pascoli, I-06123 Perugia, Italy
}

\maketitle

\begin{abstract}
We compare the high energy behaviour of hadronic photon-photon 
cross-sections in different models. We find that the photon-photon 
cross-section appears to rise faster than the purely hadronic ones
($pp$ and $p\bar{p}$). 
\end{abstract}

\section{Introduction}

\begin{figure}[h]
\centerline{\includegraphics*[width=1.0\columnwidth]{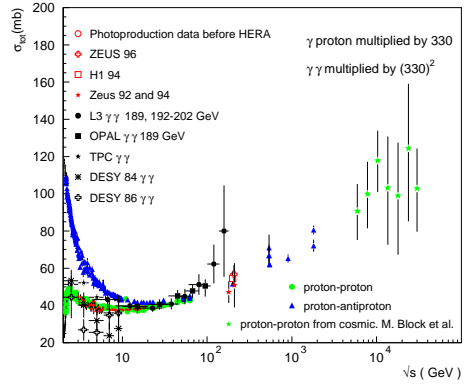}}
\caption{Proton \cite{dataproton} and photon \cite{datagp,datagg} normalized  
total cross-sections 
}\label{Fig:TX}
\end{figure}

Experimentally, all total cross-sections rise asymptotically with 
energy, but it is not clear whether the rate of increase is the same for 
different  processes.  To appreciate it  at a glance, we show in 
Fig. ~\ref{Fig:TX} a compilation of available data on $pp,\ p\bar{p}, 
\ \gamma p$ and $\gamma \gamma$ . The data span an energy 
range of four orders of magnitude. To plot them all on the same 
scale,  we have multiplied the relevant cross-section  
by a constant factor $1/R_{\gamma}$ for each incident photon. In this figure, we have taken  
$R_{\gamma}$ to be $1/330$, purely on the grounds to have all the cross-sections in the low energy region to be close in value to each other. 
One simple way to estimate this value
consists in counting the fermion 
lines in the proton and the photon and the probability of basic 
quark-antiquark scattering. Through this the factor is  
found to be 
\begin{equation}
R_{\gamma}
\approx \alpha_{QED}\
\left({{N_{fermion\ lines}^{photon}} \over {N_{fermion\ lines}^{hadron}} }\right)^2 \approx {{1}\over{300}}
\end{equation}
The same result is obtained using Vector Meson Dominance (VMD) with pure 
$\gamma-\rho$ coupling \cite{collins}.  This scaling is very approximate. 
Further there is no reason to expect 
the scaling factor to be energy independent. This can be easily 
understood by noting that at low energy the photon behaves like a vector 
meson in its interactions with matter, while at high energies QCD phenomena 
which are energy dependent will appear. Thus while at low energies
the factor $R_{\gamma}$  can be  evaluated  through VMD 
considerations~\cite{halzen,ourph01} which may include other Vector Mesons 
beyond the $\rho$,  at high energy it is likely to be different~\cite{desy}
due to the difference in the quark and gluon content of 
photons~\cite{REVPHOTSTR} and  that of the hadrons.

To  understand the role played by the quark-parton structure of the photon, 
one can use a QCD model such as the one developed in \cite{ourlast,kazimierz,
corsetti,pramana,lastPLB}, and apply it to evaluate  cross-sections for 
processes involving photons. In the following, we shall describe 
the model for protons and then apply it to photons.
\section{The Bloch-Nordsieck model for proton processes}
This model is based on the following :
\begin{enumerate}
\item QCD mini-jets to drive the rise of the total cross-section 
in the QCD asymptotic freedom regime;
\item resummation of soft gluon emission down to zero momentum  to 
soften the rise due to the increasing number of gluon-gluon collisions 
between hard perturbative, but low-x, gluons;
\item eikonal representation for the total cross-section (with $\Re e 
\chi \approx 0$) to incorporate the mini-jet cross-section, using  an impact 
parameter distribution obtained as the  Fourier transform of resummed 
soft gluon transverse momentum distribution.
\end{enumerate}
Each of these components  will be discussed in detail, before 
applying it to obtain the total cross-section.
\subsection{The mini-jet cross-section}
The mini-jet cross-section is obtained by integrating the standard QCD 
inclusive jet cross-section, using  a lower cutoff $p_{tmin}$
$ \approx 1\ GeV$, and is given by:  
\begin{eqnarray}
\sigma_{hard} \equiv \sigma^{AB}_{\rm jet} (s,p_{tmin})= \nonumber\\
\int_{p_{tmin}}^{\sqrt{s}/2} d p_t \int_{4
p_t^2/s}^1 d x_1  \int_{4 p_t^2/(x_1 s)}^1 d x_2 \nonumber \\
\times
\sum_{i,j,k,l}
f_{i|A}(x_1,p_t^2) f_{j|B}(x_2,p_t^2)
  \frac { d \hat{\sigma}_{ij}^{ kl}(\hat{s})} {d p_t}
  \end{eqnarray}
where $A$ and $B$ are the colliding hadrons. This cross-section strongly depends on the value 
chosen for such $p_{tmin}$ and -for a fixed cut off- it increases with
energy, reflecting the sharp increase in the number of low-x gluons with 
increasing energy.  This increase is very rapid and, if left unchecked, the 
mini-jet cross-section would surpass the observed total cross-section. The 
saturation mechanism which restores the Froissart bound comes from another 
QCD mechanism, soft gluon emission, which will be described in the next 
subsection.

\begin{figure}[h]
\centerline{\includegraphics*[width=1.0\columnwidth]{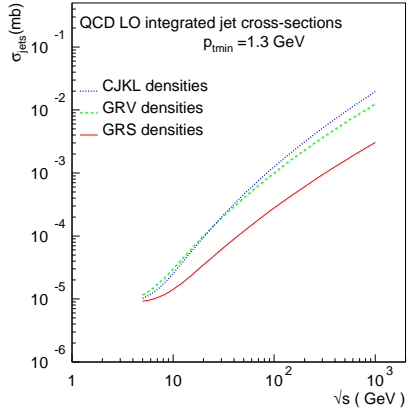}}
\caption{Photon-photon jet cross-sections for different densities 
and  a typical $p_{tmin}$ value.
}\label{Fig:miniphoton}
\end{figure}

The mini-jet cross-sections are calculated using realistic parton 
densities (PDFs). The most common ones for the proton are 
GRV \cite{GRV},   MRST \cite{MRST}, CTEQ \cite{CTEQ}, whereas 
those for the photon are GRV\cite{GRVPHO}, GRS \cite{GRS}, CJKL \cite{ CJKL}. 
These densities are available both at leading order (LO) or 
higher, but in our model we use only the LO given the fact that
 part of the NLO effects are described by the soft gluon 
resummation discussed in the next section. 
We show in Figure \ref{Fig:miniphoton} the energy dependence of the
mini-jet cross-sections for $\gamma \gamma$ collisions,
for three different  sets of parton densities, GRV, GRS and CJKL,  
for a typical  value of the cut-off, $p_{tmin}=1.3\  GeV$.
\subsection{Soft Gluons and the infrared limit}
We stress the distinction between low-x gluons which  participate in hard 
parton-parton scattering described by the mini-jet 
cross-section discussed in the previous section, and the soft gluons 
emitted in any given parton-parton process. Soft gluons by definition need 
to be resummed, and hence their momentum integrated up to a maximum value. 
The maximum value should typically be $10\div 20 \%
$ of the emitting parton energy, the lower value is usually taken to 
correspond to the intrinsic transverse momentum scale of the scattering 
hadron. Instead we extend the integration down to the zero momentum modes. 
To do so, we need therefore to make an ans\"atz as to  the behaviour of the 
strong coupling constant in the infrared region,
where the 
usual asymptotic freedom expression for $\alpha_s(Q^2)$ cannot be used. 
Our proposal is that in the infrared limit, one can phenomenologically 
use the expression
\begin{equation}
\alpha_s(k_t)= constant \times \left({{\Lambda}\over{k_t}}\right)^{2p}\ \ \ \ k_t\to 0
\label{phenoas}
\end{equation}
where $\Lambda$ is the QCD constant in the scheme chosen for the hard 
scattering calculation, and $p$ is a parameter which embeds the 
infrared behavior with  $p<1$ so that the soft gluon integrals 
converge. The constant in front of Eq.~\ref{phenoas} should be chosen to 
provide a smooth extrapolation to the perturbative expression for 
$\alpha_s$. Our choice for the interpolating function is 
\begin{equation}
\alpha_s={{12 \pi}\over{33-2N_f}} {{p}
\over{\ln[1+p({{k_t}\over{\Lambda}})^{2p}]}}
\label{alphas}
\end{equation}
This expression is used in the soft gluon resummation formula in the 
transverse momentum variable which reads:
\begin{equation}
d^2P({\bf K}_\perp)=d^2 {\bf K}_\perp 
\int {{d^2 {\bf b}} \over {(2 \pi)^2}} 
e^{i{\bf K_\perp\cdot b} -h( b,q_{max})}
\end{equation}
with
\begin{equation}
h( b,q_{max}) =\int_0^{q_{max}}  d^3{\bar n}(k) [1-e^{-{\bf k_t\cdot b}}]
\end{equation}
In QED $d^3{\bar n}(k)\propto \alpha \log ( {{ 2q_{max} } \over{
m_{electron} }})$ and resummation in transverse momentum variable
is well approximated by   first order expansion in $\alpha$. In
QCD, the situation is completely different because $\alpha_s$ is
(i) not a constant and (ii)  can become very large as the gluon
transverse momentum goes to zero. In phenomenological
applications, resummation is typically exploited by splitting the
integral between the very soft and the non-infrared region so that
\begin{eqnarray}
&h(b,q_{max}) =c_0 b^2 + \int_\mu^{q_{max}}  d^3{\bar n}(k)
[1-e^{-i{\bf k_t\cdot b}}] \nonumber \\
& \approx c_0 b^2 + c_1 \int_\mu^{q_{max}}
{{dk_t^2}\over{k_t^2}} \alpha_s(k_t^2) \log
({{2q_{max}}\over{k_t}})
\end{eqnarray}
As mentioned before, our approach is different. We use Eq. \ref{alphas}
and 
\begin{eqnarray}
&h( b,q_{max})  = 
\frac{16}{3}\int_0^{q_{max} }
{{dk_t}\over{k_t}}  {{ \alpha_s(k_t^2) }\over{\pi}}      \nonumber \\
&~~~\left(\log{{2q_{\max}}\over{k_t}}\right)\left[1-J_0(k_tb)\right].
\end{eqnarray}
The energy dependent quantity $q_{\max}$ represents the 
maximum energy allowed to a soft gluon in a given parton-parton 
interaction, and it depends upon the kinematics, i.e. 
the parton energy fractions $x_1$ and $x_2$, and from the emitted parton 
momentum $p_t$. (This is described explicitly in \cite{kazimierz} ). 
\subsection{Embedding QCD in the eikonal}
\begin{figure}[h]
\centerline{\includegraphics*[width=1.0\columnwidth]{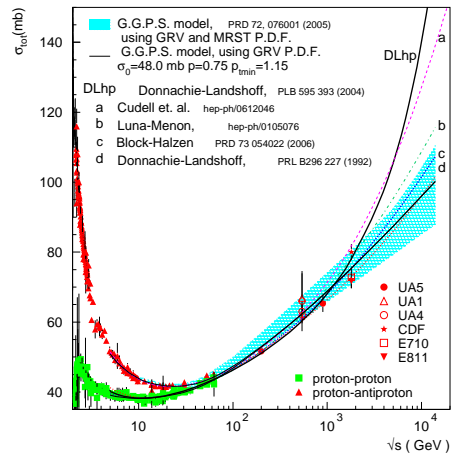}}
\caption{Proton-proton and proton-antiproton  total cross-sections from the 
Bloch-Nordsieck model   described in the text. A discussion of
comparison with other models and explanation of symbols can be found in 
\protect\cite{lastPLB}.}\label{Fig:regge}
\end{figure}

A convenient formalism to calculate total cross-sections is provided by the 
eikonal integral for the elastic amplitude, valid for small angle
scattering. Using the optical theorem, the total cross-section can be
written as
\begin{equation}
\sigma_{tot}=2\int d^2{\bf b}[1-e^{-\Im m \chi(b,s)}\cos \Re e \chi(b,s)]
\end{equation}
A simple model for calculating the eikonal function $\chi(b,s)$ consists in 
evaluating the probability of inelastic processes, obtained by summing over 
all possible Poisson distributed collisions. One then obtains
\begin{equation}
\sigma_{inel}=\int d^2{\bf b}[1-e^{-n(b,s)}]
\end{equation}
where $n(b,s)$ is the average number of inelastic collisions. 
Neglecting the real part of the eikonal, one finds
\begin{equation}
\sigma_{tot}=2\int d^2{\bf b}[1-e^{-n(b,s)/2}]
\end{equation}
This expression has the advantage of satisfying unitarity, but it requires 
knowledge of the impact parameter distribution of the scattering
particles. 
We propose this distribution to be given by the Fourier transform of the 
soft gluon distribution discussed in the previous section, namely
\begin{eqnarray}
 A_{BN}(b,s) =
N \int d^2 {\bf K}_{\perp} {{d^2P({\bf K}_\perp)}\over{d^2 {\bf K}_\perp}} 
 e^{-i{\bf K}_\perp\cdot {\bf b}} \nonumber \\
 = {{e^{-h( b,q_{max})}}\over
 {\int d^2{\bf b} e^{-h(b,q_{max})}
 }}
 \label{abn}
 \end{eqnarray}
 and  approximate $n(b,s)$ as
 \begin{eqnarray}
&n(b,s)=n_{soft}(b,s) +
n_{hard}(b,s) =   \nonumber \\
&n_{soft}(b,s)+A_{BN}(b,s)\sigma_{jet}(s,p_{tmin})
 \end{eqnarray}
 where $n_{hard}(b,s)$ represents the average number of collisions with 
outgoing partons with $p_t>p_{tmin}$, with all other collisions of non 
perturbative description included in $n_{soft}(b,s)$. This quantity is 
often parametrized by factorizing completely the $b-$ and $s-$ dependence, 
and using the Fourier transform of the relevant hadron form factor to 
describe the impact parameter distribution. 
Instead, in our model, we do not use a factorised form in $b$ and $s$,
and write
 \begin{equation}
 n_{soft}(b,s)=A_{BN}^{soft}(b,s)\sigma_0
 [
 1+\epsilon{{2}\over{\sqrt{s}
 }}
 ]
 \end{equation}
 with $\sigma_0$ a constant to fix the overall normalization, and 
$\epsilon=0,1$ 
depending upon the process being proton-proton or proton-antiproton 
respectively.  Here, $A_{BN}^{soft}(b,s)$ is 
obtained from the Bloch-Nordsieck (BN) model 
with $q_{max}^{soft}$ parametrized 
so as to always remain $< 10\div 20 \%$ of the value of $p_{tmin}$.
For proton-proton and proton-antiproton scattering we show the results of 
our model  in Figure \ref{Fig:regge}, where we also show comparison with 
other models \cite{bibicross}. 
\section{Photon processes}
\begin{figure}[h]
\centerline{\includegraphics*[width=1.0\columnwidth]{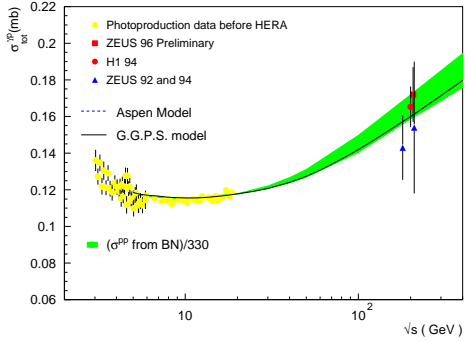}}
\caption{Total photon-proton  cross-sections with  the full line corresponding 
to  GGPS model full line of Fig.  \ref{Fig:regge} and the dotted line 
to \cite{aspen}.}\label{Fig:gamp}
\end{figure}

\begin{figure}[h]
\centerline{
\includegraphics*[width=1.0\columnwidth]{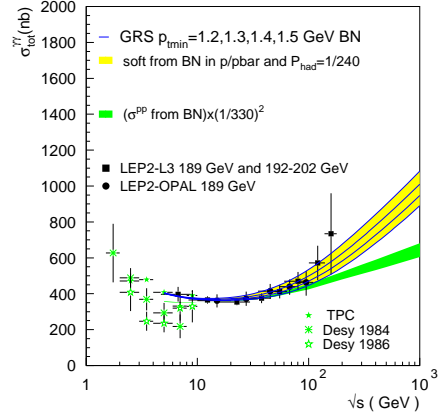}}
\caption{
$\gamma \gamma $ total cross-sections from factorization
(green band) or using the Eikonal  with  GRS densities  and   soft gluon 
resummation.}
\label{Fig:gg}
\end{figure}

The   model described previously can now  be applied to the photon processes. 
An extremely simple exercise is to use factorization as 
\begin{equation}
\sigma_{\gamma \gamma}=
{
{
(\sigma_{\gamma p})^2}
\over{
\sigma_{pp}}}.
\label{Eq:gribov}
\end{equation}
One can then either parametrize  $\sigma_{\gamma p}$ or choose  an 
appropriate constant as we did in Fig. \ref{Fig:TX}.
The result is shown in Figs. \ref{Fig:gamp} and \ref{Fig:gg} where the green band 
corresponds to
the band for proton-proton scattering of Fig. \ref{Fig:regge} multiplied by $1/330$ and $(1/330)^2 $ respectively. 

A more involved exercise is  to follow the same eikonalization procedure as
outlined for the proton cross-sections, using the experimentally 
determined  photonic parton densities,  and then compare it with data,
as well as other model predictions.  In such case, one needs to know how to 
apply to the photon a typically hadronic description like the eikonal 
representation. Quite a while ago \cite{halzen}, the following expression 
was proposed for photon processes:
\begin{equation}
\sigma_{tot}=2[P_{had}]^l\int d^2{\bf b}[1-e^{-n^{\gamma}(b,s)/2}]
\label{Eq:eikg}
\end{equation}
with
\begin{equation}
n^{\gamma}(b,s)=[{{2}\over{3}}]^{l}n_{soft}(b,s)+n_{hard}^{\gamma}
\label{Eq:nb}
\end{equation}
and with
\begin{equation}
n_{hard}^{\gamma}=A_{BN}^{\gamma}(b,s)
{{
\sigma_{jet}(s,p_{tmin})}\over{[P_{had}]^l}}
\label{Eq:nhard}
\end{equation}
with $l=1,2$ for $\gamma p$ and $\gamma \gamma$ respectively, 
$\sigma_{jet}(s,p_{tmin})$ to be calculated using current photon 
densities, and $A_{BN}^{\gamma}(b,s)$ given by Eq.\ref{abn} with the 
appropriate $q_{max}$ for the photon processes.

In the above described expressions, there appears the quantity $P_{had}$, 
 which represents the probability 
that the photon behaves like a hadron in the eikonal formulation.
A possibility is to use VMD models, namely
\begin{equation}
P_{had}=P_{VMD}=\sum_{V=\rho,\omega,\phi} {{4\pi \alpha}
\over{f^2_V}}= {{1}\over{250}}
\label{Eq:VMD}
\end{equation}
where the sum extends to all  vector mesons, not just the $\rho$.
As the photon energy increases, the contribution from
the other, resolved, components also increases and one can expect 
$P_{had}$ to differ from the VMD proposal. 
Using a phenomenologically 
fixed value $P_{had}=1/240$
we 
obtain the results  shown in Fig. \ref{Fig:gg} for the $\gamma \gamma$ case, using GRS densities. 

Please notice that the parameter $R_\gamma$ appearing elsewhere in this note
is a purely phenomenological multiplicative factor between the total
photon and proton cross-sections, and it is different from $P_{had}$, which
is defined through the eikonal and $not$ through the cross-sections. 
It is however to be expected that both parameters be of the same order of
magnitude, namely of $O (\alpha)$.
\section{Conclusions}
We notice that while factorization a' la Gribov could still hold for $\gamma p$,
the same cannot be said for $\gamma \gamma$ where present data do appear to 
be higher than the curve obtained by the simple factorization hypothesis. On the other hand a more detailed model, like the Mini-jet cum soft gluon resummation appears better suited to describe the present photon-photon data.

\section{Acknowledgments}
 G. P.  thanks  the organizers  of Photon2007 for the hospitality
and the very beautiful venue of the Conference. 
This work has been partially supported by MEC (FPA2006-05294) and by Junta
de 
Andaluc\'\i a (FQM 101 and FQM 437).

\begin{footnotesize}

\end{footnotesize}
\end{document}